\documentclass[12pt]{article}
\newenvironment{makeimage}{}{}
\newcommand{\latex}[1]{#1}
\newcommand{\itemhead}[1]{\par\medskip\noindent{\bf #1}}
\newcommand{\mod}[1]{\;(\mathop{\mbox{\rm mod}} #1)}

\newtheorem{exercise}{Exercise}

\title{Introduction to Public Key Cryptography and Modular Arithmetic.}
\author{Robert Milson}
\begin{document}
\maketitle
\section{Introduction.}

These notes are an introduction to the RSA algorithm, and to the
mathematics needed to understand it.  The RSA algorithm --- the name
comes from the initials of its inventors, Rivest, Shamir, and Adleman
--- is the foundation of modern public key cryptography.  It is used
for electronic commerce and many other types of secure communication
over the Internet.

The RSA algorithm is based on a type of mathematics known as modular
arithmetic.  Modular arithmetic is an interesting variation of
ordinary arithmetic, but whereas everyday arithmetic is familiar to
school children everywhere, modular arithmetic is a somewhat obscure
subject.  It's not that modular arithmetic is particularly difficult,
or confusing; one could teach it in high school, or even earlier.
Conventional thinking, however, places a higher value on the ability
to balance a checkbook than on the ability to communicate in code;
this is probably the reason why most people have never heard of
modular arithmetic.  Today, the rapid proliferation of the Internet
and the growing popularity of electronic financial transactions are
causing a shift in attitudes.  These days, even an introductory
understanding of public key cryptography can be enormously useful.
Consequently, today's students deserve an opportunity to become
acquainted with the methods and ideas of modular arithmetic.

First, a quick word about these notes.  I have tried to make the
material here as down to earth, and accessible as possible.  As such,
the emphasis is on concrete calculations, rather than abstruse theory.
My goal is to guide you through the concrete steps needed to implement
and understand RSA encryption/decryption.  Therefore in the present
context, calculations are in and abstraction is out.  A number of
simple exercises is included; you should work them as you go along.
The answers to the exercises are collected in a final appendix.

So set aside a few hours, go grab a pencil, and pour yourself a cup of
coffee.  It shouldn't take very long to go through these notes; the
material in here just isn't all that hard.

\section{Modular Arithmetic} 

I remember that as a child, after I learned the general system of
counting, I was fascinated by and frequently thought about the fact
that {\em numbers never end.}  In principle, one can count as high as
one wants --- an activity I experimented with as a child, and one that
serves me well in adulthood when insomnia comes calling.  What,
however, would be the consequences if numbers {\em did end}?  What if
numbers behaved like the numbers on a clock?  What if, as one counted
higher and higher, one would come full circle and begin counting again
from the beginning, from zero?

To make this idea concrete suppose that there are only five different
numbers: $0, 1, 2, 3, 4$; and that these numbers are arranged on a
circular clock in the usual clock-wise direction, as in the figure
below.  It's not hard to imagine how to do addition using this
alternate system of counting.

\medskip
\begin{makeimage}
\Large\tt
\halign{&\hfil #\hfil\cr
 & &0\cr
4& & & &1\cr
 &3& &2\cr
}
\end{makeimage}
\bigskip

\noindent
Doing $2+2$ one starts with $2$, increments twice, and obtains the
answer $4$.  Doing $2+4$, however, is a little unusual, because this
time the answer is $1$.  The reason is that after $3$ steps one comes
full circle, i.e. to $0$; from there, one more increment brings the
final total to $1$.  Subtraction isn't hard to understand either; one
counts backwards, i.e. counter-clockwise, rather than forwards.  So,
in the clock-like way of counting, $2-4=3$; one starts with $2$ and
turns the clock back $4$ times.

The usual name for this kind of counting is modular arithmetic ( some
people have also called it ``clock arithmetic''.) It is important to
note that there are many different modular arithmetics; it all depends
on how big the clock is.  The size of the clock, i.e. the total of all
possible numbers, is called the {\em modulus}.  In the example of the
preceding paragraph the modulus was $5$.  One can just as happily work
with other moduli: $7$, $10$, $123$; any integer greater than $1$ will
do.  For example, the binary arithmetic of ones and zeroes that
underlies the workings of modern computers is nothing but arithmetic
with a modulus of $2$.

In addition to the clock analogy there is another explanation of
modular arithmetic that needs to be mentioned.  This alternate system
is made up of all the possible integers; however any two numbers that
generate the same displacement on a clock are considered to be
equivalent.  Thus modulo $5$ the numbers $-8, -3,2 , 7, 12, 17, 21$
are all considered to be equivalent, because they all correspond to a
displacement of $2$ spaces on a $5$-space clock.  To put it another
way, two numbers are considered equivalent if and only if they differ
by a multiple of $5$.  The mathematical notation for this equivalence
works like this.  One writes:
$$6\equiv 16 \quad\mod 5,$$
and says out loud: ``six is equivalent to
sixteen modulo five''.  The meaning of this statement is that $6$ and
$16$ differ by a multiple of $5$; or, equivalently that $6$ and $16$
generate an equivalent displacement on a clock with $5$ spaces.  In
summary, in the modulo $5$ system, there are exactly five classes of
numbers: the numbers equivalent to $0$, the numbers equivalent to $1$,
to $2$, to $3$, and to $4$.

\begin{exercise}
\label{ex:add}
Which of the following statements are true, and which are false?
{\small
$$1-8 \equiv 0 \mod7,\quad
1-8\equiv 3\mod5,\quad
9+9+9\equiv 8\mod{13},\quad
2+11-33\equiv0\mod{10}, $$}
\end{exercise}

In a system of modular arithmetic with modulus $n$ it is possible to
reduce every integer to an equivalent number between $0$ and $n-1$.
To do the reduction one divides and calculates the remainder.  For
example, let's say one wants to calculate the reduced form of $2040$
modulo $209$.  One has
$$\textstyle 2040\div209= 9 \frac{159}{209},\quad\mbox{i.e. }
2040=9\times 209+159.$$
Therefore $2040\equiv 159\mod{209}$.  
Again, the meaning of this statement is that a total displacement of
$2040$ spaces on  a clock with $209$ places yields a net displacement
of $159$ spaces. 

By the way, save yourself some grief and check the documentation of
that expensive calculator you bought when you came to university.
Many modern calculators have a built-in remainder function, and
possibly other functions for performing the operations of modular
arithmetic.

\subsection{Multiplication}
As everyone knows, multiplication is nothing but repeated addition.
This is the meaning of multiplication in everyday arithmetic; this is
how multiplication works in modular arithmetic as well.  Say one wants
to calculate $2\times 3$, i.e.  $2+2+2$, in the modulo $5$ system.
The answer of course is $6$, which modulo $5$ is equivalent to $1$.
Writing this more succinctly: $2\times3 \equiv 1 \mod5$.

How sensible is the above definition of multiplication?  Does it obey
the same algebraic laws as ordinary multiplication?  The answer is,
yes!  For example, ordinary multiplication obeys a rule called the
distributive law.  As a particular instance of this rule, $(3+4)\times
2=7\times 2$ is equal to $3\times 2 +4\times 2=6+8$.  

What about modular arithmetic; does the distributive law continue to
work?  Consider the last calculations modulo $5$.  For the first
expression one gets
$$(3+4)\times 2=7\times 2 \equiv 2\times 2\equiv 4\mod 5.$$
For the
second expression one gets
$$3\times2 + 4\times 2=6+8\equiv 1+3\equiv 4\mod 5;$$
the two answers
agree.

What about something like $8\times 2$ versus $3\times 2$?  Since $3$
and $8$ are equivalent modulo $5$, it stands to reason that the two
answers should also be equivalent.  This is indeed the case; one gets
an answer of $1 \mod 5$ for both calculations.  The answers are the
same because $8$ and $3$ differ by a multiple of $5$; indeed, $8=3+5$.
Therefore 
$$2\times 8 = 2\times(3+5) = 2\times3+2\times 5.$$
Now $2\times 5$ is
a multiple of $5$ and is consequently equivalent to $0 \mod 5$.
Therefore
$$2\times 8 = 2\times3 + 2\times 5\equiv 2\times 3 + 0 \mod 5.$$
In
summary, multiplication is a perfectly valid, consistent operation in
the world of modular arithmetic.

\begin{exercise}
\label{ex:mtable}
Complete the following table of multiplication modulo $5$.

\medskip
\begin{makeimage}
\large
\begin{tabular}{c|cccc}
$\times$ &  $1$ & $2$ & $3$ & $4$ \\
\hline
$1$ & $1$ & $2$ & $3$ & $4$  \\
$2$ & $2$ & $4$ & $1$ & $3$  \\
$3$ & . & . & . & .  \\
$4$ & . & . & . & .  \\
\end{tabular}
\end{makeimage}

\end{exercise}

A surprising aspect of modular arithmetic is the fact that one can
also do division.  Recall that every division operation in ordinary
arithmetic can be recast as a corresponding multiplication by a
reciprocal.  Thus, $a\div b= a\times b^{-1}$, where $b^{-1}$ is the
one particular number such that $b\times b^{-1}=1$.  A moment's worth
of reflection will reveal that this definition makes perfect sense for
modular arithmetic.  Say one wanted to  find the reciprocal of $2$
modulo $5$.  A consultation of the multiplication table from the
preceding exercise readily yields the answer: $2\times 3\equiv 1\mod
5$, and therefore one writes $2^{-1}\equiv 3 \mod 5$.  In a sense,
division in modular arithmetic is easier than it is in ordinary
arithmetic; one doesn't have to worry about fractions.  For example,
division by $2$ is equivalent to a multiplication by $3$.  Thus
$3\div2\equiv 3\times 3\equiv 4\mod 5$.  One can even check the
answer: $4\times2\equiv 3\mod5$; everything is correct.
\begin{exercise}
\label{ex:reciprocal}
Find the reciprocals of $3$ and $4$ modulo $5$.  Perform the following
divisions modulo $5$ and then check the answers by performing the
necessary multiplications: $4\div 3$, $3\div 4$, $3\div 3$, $1\div 3$.
\end{exercise}

\subsection{It's a strange world, after all.}
Who can forget that old bugaboo of elementary school arithmetic: the
problem of division by zero?  ``Division by zero is ...'' There are a
number of typical endings to this sentence, and all of them serve to
imply that division by zero is somehow bad, if not outright
impossible.  My preference in the face of such a question is to be as
undogmatic as possible, and to pass the ball back to the questioner:
``I don't know what one divided by zero is, my friend, but if you
would like to hazard an answer, I will be happy to check it for you.
What did you say?  One divided by zero is three?  I don't think so.
You see, zero times three is zero, not one; so I'm afraid you shall
have to try again.''  In this fashion the questioner should quickly
become convinced that $1\div 0$, whatever such an answer may be,
cannot be an ordinary, everyday number.  So perhaps what one should
really be saying is that the equation $\mbox{?}\times 0=1$ has no
solutions.

Division by zero is just as problematic in modular arithmetic as it is
in ordinary arithmetic.  Furthermore, in the world of modular
arithmetic, there exist, in addition to zero, other ``division
unfriendly'' numbers.  In order to see this, consider the following
operation in the modulo $6$ system: $1\div 4$.  In other words, try to
solve the following equation: $4\times \mbox{?}\equiv1\mod6$.  No such
solution exists, of course.  Think about the multiplication table
modulo $6$.  The row that begins with $4$ reads: $4, 2, 0 , 4 , 2$.
In contrast to the way multiplication worked modulo $5$, certain rows
of the multiplication table modulo $6$ have repeated entries, and
consequently, in such rows, certain other numbers do not appear at
all.  So $1\div 4$ cannot be given a value, because the number $1$
doesn't occur in row number $4$.  Notice that something like $4\div 4$
doesn't work either, but now ambiguity is to blame.  There are two
possible solutions to the equation $4\times x\equiv 4\mod 6$; both
$x=1$ and $x=4$ will work.

\begin{exercise}
\label{ex:mtable6}
Write down the multiplication table modulo $6$.
\end{exercise}

These strange ``division unfriendly'' numbers possess another curious
property.  Notice that $4\times 3\equiv 0\mod 6$.  One would never
expect to see an equation like that in ordinary arithmetic; one simply
doesn't expect to multiply two non-zero numbers and have the answer
turn out to be zero.  As unusual as this may appear at first glance,
such goings on are quite commonplace in the world of modular
arithmetic.  In fact, there is some standard terminology that serves
to describe such situations.  If the product of two given numbers is
zero, one calls these numbers {\em divisors of zero.}  It's an apt
title, because such numbers are literally able to divide zero and have
a non-zero number be the answer.   Modulo $6$ the
divisors of zero are $2$, $3$, and $4$, because
$$2\times 3 \equiv 4\times 3 \equiv 0 \mod 6.$$

What about the remaining numbers in the mod $6$ system; what about $1$
and $5$?  These are the ``division friendly'' numbers.  Again there is
some standard terminology that one should learn at this point.  A
number is called {\em a unit} if and only if it possesses a
multiplicative reciprocal.  Thus, both $1$ and $5$ have reciprocals
and are therefore called units: $1^{-1}\equiv 1\mod6$ and
$5^{-1}\equiv 5\mod 6$

\begin{exercise}
\label{ex:unit}
Explain why in modular arithmetic a unit can never be a divisor of
zero, and why a divisor of zero can never be a unit.
\end{exercise}

\subsection{GCD: the Greatest Common Divisor}
The acronym GCD stands for {\em greatest common divisor}.  A common
divisor of numbers $a$ and $b$ is a number that evenly divides both
$a$ and $b$.  The greatest common divisor of $a$ and $b$ is simply the
largest such common divisor.  Consider for example the numbers $12$
and $16$.  Obviously $2$ is a common divisor of both.  However, $4$ is
also a common divisor, and in fact it is the largest common divisor.
Therefore $4$ is the GCD of $12$ and $16$.

Note that $1$ is a common divisor of every possible pair of numbers.
If $1$ is the {\em only} common divisor of $a$ and $b$, then one says
that $a$ and $b$ are {\em relatively prime}.  For example, $9$ and
$16$ are relatively prime.

\begin{exercise}
\label{ex:gcd1}
Calculate the GCD of the following pairs of numbers:  $6$ and $8$; $6$
and $18$; $7$ and $15$; $30$ and $20$.
\end{exercise}

Common divisors are a very important idea in modular
arithmetic; they are the key concept for understanding
divisors of zero.
\begin{quote}
  \itemhead{Fact.}  If numbers $x$ and $n$ have a common divisor other
  than $1$, then $x$ is a divisor of zero modulo $n$.
\end{quote}
\noindent
For example, $2$ is a common divisor of $6$ and $8$, and consequently
$6$ is a divisor of zero, modulo $8$.  Indeed,
$6\times4=24\equiv0\mod8$.  It isn't difficult to explain this fact.
Let $d>1$ be a common divisor of $x$ and $n$.  By definition, $x\div
d$ and $n\div d$ are both whole numbers, and therefore
$$x\times(n\div d) = (x\div d)\times n \equiv
0\mod{n}.$$

\begin{exercise}
\label{ex:10sys}
Write down the multiplication table modulo $10$.  Identify the units
and the divisors of zero.  Notice that $10=2\times 5$.  How is this
fact relevant to your search for units and divisors of zero?
\end{exercise}

\subsection{The Euclidean Algorithm}
It is easy enough to guess a GCD of two small numbers, but larger
numbers require a more systematic approach.  Fortunately there is a
straightforward method, called the Euclidean algorithm, for
calculating the GCD of two whole numbers.  The algorithm is based on
the following idea.  Suppose the goal is to find the GCD of a pair of
whole numbers $x$ and $y$. It isn't hard to see that if a number $d$
evenly divides both $x$ and $y$, then $d$ will also divide all of the
following numbers: $x-y$, $x-2y$, $x-3y$, etc. Indeed $d$ will evenly
divide any number of the form $x-p\times y$.  Conversely if $d$ evenly
divides $y$ and $x-p\times y$, then $d$ will also evenly divide $x$.
\begin{quote}
\itemhead{Conclusion:} for every whole number $p$, the
GCD of $x$ and $y$ is equal to the GCD of $y$ and $x-p\times y$.
\end{quote}
For example, say one wants to
compute the GCD of $168$ and $91$.  One may as well be looking for
the GCD of $91$ and $168-91=77$.  Repeating this reasoning, one
should next look for the GCD of $77$ and $91-77=14$, and then
for the GCD of $14$ and $77-5\times 14=7$.  Now $14=2\times 7$, and
therefore it is clear that $7$ is the desired GCD.  Indeed: $168\div
7=24$ and $91\div 7=13$.  The algorithm is summarized below.
\begin{quote}
  \itemhead{The Euclidean Algorithm.}  Goal: to find the GCD of whole
  numbers $x>y$.  If $y$ divides $x$ evenly then $y$ is the GCD.
  Otherwise, calculate the remainder, $r$, from the division of $x$ by
  $y$.  In other words, $r=x-p\times y$ where $p$ is the whole part of
  $x\div y$.  Since GCD($x,y$)=GCD($y,r$), one goes back to the
  beginning of the algorithm and restarts the calculation with the new
  $x$ equal to the old $y$, and the new $y$ equal to the old $r$.
  Since $y<x$ and $r<y$, the inputs to the algorithm become smaller
  with each iteration. The algorithm is, therefore, guaranteed to
  terminate after a finite number of repetitions.
\end{quote}

\begin{exercise}
\label{ex:gcd2}
Use the Euclidean algorithm to find the GCD of $1113$ and $504$.
\end{exercise}

It is easy enough to see that if $d$ evenly divides a pair of numbers
$x$ and $y$, then $d$ will also evenly divide $a\times x+b\times y$
for all integers $a$ and $b$.  The following fact is a little more
surprising.
\begin{quote}
\itemhead{Fact.}  Given whole numbers $x$ and $y$, one can find
integers $a$ and $b$ so that $a\times x+b\times y$ is exactly equal to
the GCD of $x$ and of $y$.  In particular, if $x$ and $y$ are
relatively prime, then one can find $a$ and $b$ so that $a\times
x+b\times y=1$.
\end{quote}
This fact is enormously useful in connection with the calculation of
reciprocals in modular arithmetic; this point will be explained
shortly.  First, however, one needs to master a modified version of
the Euclidean algorithm, a version tailored to the calculation of the
critical $a$ and $b$.  Proceeding by way of example, suppose one is
interested in the GCD of $1113$ and $504$.  Consider the following
table.

\medskip
\begin{makeimage}
\hskip 1em\vbox{\offinterlineskip
{\halign{\tabskip 2pt\hfil \quad\strut $#$  && \vrule\hfil\quad $#$ \cr
n    &p & a & b \cr
\noalign{\hrule}
1113 &  & 1 & 0  \cr
504  &2 & 0 & 1  \cr
105  &4 & 1 &-2  \cr
84   &1 &-4 & 9  \cr
21   &4 & 5 &-11 \cr
0    & \cr
}}}
\end{makeimage}
\medskip

In what follows, subscripts are used to specify the number of the row.
Thus $n_3$ refers to $105$, the $n$ entry in the third row.  The first
column contains the sequence of numbers obtained in the course of
applying the Euclidean algorithm (c.f. Exercise \ref{ex:gcd2}).  In
other words, $n_3$ is the remainder of $n_1\div n_2$, $n_4$ is the
remainder of $n_2\div n_3$, etc.  The $p$ column contains the whole
part of each of these divisions: $p_2$ is the whole part of $n_1\div
n_2$, $p_3$ is the whole part of $n_2\div n_3$, etc.  In other words,

\vskip -10pt
\begin{makeimage}
\begin{eqnarray*}
n_3&=&n_1-p_2\times n_2,\\
n_4&=&n_2-p_3\times n_3,\\
n_5&=&n_3-p_4\times n_4,
\end{eqnarray*}
\end{makeimage}
\vskip -10pt

The entries in the $a$ and $b$ columns are chosen so that $n=a\times
1113+b\times 504$ in each row of the table.  For example, row four
contains $84=-4\times 1113+9\times 504$.  The final row (not counting
the last zero) contains the final, desired $a$ and $b$.  Indeed,
$21=5\times 1113-11\times 504$.  For obvious reasons, $a=1$, $b=0$ in
the first row, and $a=0$, $b=1$ in the second row.  For the subsequent
rows the entries are calculated in the same fashion as the entries in
the $n$ column.  To be more specific:

\vskip -10pt
\begin{makeimage}
\begin{eqnarray*}
a_3&=&a_1-p_2\times a_2,\\
a_4&=&a_2-p_3\times a_3,\\
a_5&=&a_3-p_4\times a_4,
\end{eqnarray*}
\end{makeimage}
\vskip -10pt

\noindent
and so on.  The $b$ column is computed in the same way.  I will leave
it to you to figure out why this method ensures that $n=a\times
1113+b\times 504$ in every row.

\begin{exercise}
\label{ex:gcd3}
Calculate the GCD of $1466$ and $237$, and find the integers $a$, $b$
such that $a\times 1466+b\times 237$ is equal to the GCD.
\end{exercise}

You are now in possession of a method that makes it easy to calculate
reciprocals in modular arithmetic.  Remember, that $x$ is a unit
modulo $n$ if and only if $x$ and $n$ are relatively prime.  If this
is the case, then one simply finds $a$ and $b$ so that $a\times
x+b\times n=1$, and voila: $a\times x\equiv 1\mod n$, i.e. $a\equiv
x^{-1}\mod n$.

\begin{exercise}
\label{ex:gcd4}
Calculate $59\div 237$ modulo $1466$.  Check your answer.
\end{exercise}

\subsection{Curiouser and curiouser .... }
\label{sect:cc}

Presented for your reading pleasure, the following curious tidbit from
the world of recreational mathematics.  Take out a calculator and
randomly punch in a whole number.  Note the last digit of your chosen
number, and then raise your number to the fifth power.  If your
starting number wasn't too large, and if your calculator has
sufficiently many digits on its display, then you will notice that the
last digit of the result is the same as the last digit of the starting
number.  For example: $17^5 = 1419857$ and $22^5 = 5153632$.  This
interesting phenomenon occurs for other powers as well: the powers $1,
5, 9, 13, 17, 21, $ etc, will all work.  For example, $7^9 = 40353607$
and $2^{13} = 8192$.

You may have realized that the business of looking at the last digit
can be handled most conveniently in terms of modular arithmetic.  The
fact is that modulo 10, a number is equivalent to its last digit, and
therefore this curious fact can be written down as the following
identity:
$$x^p\equiv x\mod{10},\quad\mbox{as long as }
p=1,5,9,13,17,21,25,\ldots$$

The above identity is far from being a mere curiosity.  It is, in
fact, the essential component of the RSA encryption/decryption
process.  The application to cryptography will be explained in the
next section.  First, it will be necessary to take a closer look at
the above ``curious fact'', and try to find the analogous trick for
moduli other than $10$.

As you may have noticed, the critical exponents, i.e. the values of
$p$ for which $x^p\equiv x\mod{10}$, are precisely the numbers $p$
such that $p-1$ is a multiple of $4$.  To put it differently, the list
of critical exponents is obtained by starting with $1$ and then
repeatedly adding $4$.  Why the number $4$ though?  Look back at
Exercise \ref{ex:10sys}, and {\em count the number of units} in that
particular system of modular arithmetic.  That's right; there are $4$
units, and for reasons that we won't get into here, the spacing in the
list of critical exponents is always equal to the total number of
units in one's chosen system of modular arithmetic.  This
all-important principle is summarized below.
\begin{quote}
  \itemhead{A Curious Fact.}  Suppose that $n$ is a square free whole
  number, and let $\phi$ be the total number of units in the system of
  arithmetic modulo $n$.  In such circumstances the following identity
  holds: $x^p\equiv x \mod{n}$ where the exponent can be any of the
  critical values $p=1,1+\phi,1+2\phi,1+3\phi,\ldots$
\end{quote}
The business about square-free numbers is an important, but technical
detail.  The problem is that the above ``curious fact'' does not work
for certain values of the modulus $n$.  The values of $n$ for which
the trick fails to work are precisely those whole numbers that can be
divided evenly by a square.  For example, working modulo $8$ and
starting with $2$ one has
$$2^2\equiv 4,\quad 2^3\equiv 0,\quad 2^4\equiv 0,\quad 2^5\equiv
0,\quad \mbox{etc}$$
and so one never gets back to $2$, no matter
which power is used.  In other words, the trick doesn't work when $8$
is the modulus.  Now $8$ can be divided by $4$, and the latter is a
square, and that is why the trick fails to work modulo $8$.  The
bottom line is that the ``curious fact'' will work if and only if one
chooses a clock size (i.e. a modulus $n$) that does not have any
squares as a factor.

\begin{exercise}
\label{ex:22sys}
What are the critical exponents $p$ in the system of arithmetic modulo
$22$?  Do some calculations and verify that $x^p\equiv x\mod{22}$ for
these critical values of $p$.  
\end{exercise}

\begin{exercise}
\label{ex:22sysa}
Explain why, in the modulo $22$ system, it suffices to consider only
the exponents from $1$ to $10$.  Calculate $3^{32} \mod{22}$ just ``by
looking at it''.
\end{exercise}

\section{RSA}
Those of you who navigate the World Wide Web using Netscape browsers
may have noticed that the very first screen shown by Navigator and
Communicator has a bunch of corporate looking logos, including one
that has a pair of keys alongside the letters RSA.  The icon in
question is the logo of {\bf RSA Data Security}, the owners of the
patented RSA public key encryption algorithm.  The letters RSA stand
for Rivest, Shamir and Adleman; these are the names of the M.I.T.
(Massachusetts Institute of Technology) academics who invented the
algorithm in 1977.  In 1982 these three individuals went on to found
{\bf RSA Data Security}, a company that specializes in secure digital
communication.  Another significant acronym in the world of computer
cryptography is {\bf PGP}, which stands for {\em pretty good privacy}.
This is a loosely organized collection of (mostly) free software that
puts the power of private and secure digital communication in the
hands of ordinary citizens. PGP relies on the RSA algorithm for its
core functionality.

\subsection{Public key cryptography.}
What is the basic goal of cryptography? Speaking generally, one wants
a method for encoding human-readable messages into an unreadable
cipher, as well as a corresponding method for translating the coded
information back into everyday language.  Typically, a cryptographic
method involves some sort of a secret key.  One simple example is the
so-called substitution cipher.  One assigns a number to each letter of
the alphabet, and then encodes and decodes messages using this code.
In theory, people who are not privy to the translation table of
letter-number pairs, should not be able to decode and understand the
message ( A skilled codebreaker, however, can break a substitution
cipher without much trouble. )

The substitution cipher is a basic example of traditional, single key
cryptography.  The reason for the name ``single key'' ( ``secret key''
is also used ) is that both the sender and receiver of an encoded
message make use of the same, secret key --- in this case the table of
letter-number pairs --- to generate and decipher the message.  All
single key encryption schemes share a fundamental difficulty: the
encryption key must be agreed on beforehand and then kept absolutely
secret.  Thus, single key encryption is both inconvenient (one has to
find a secure means of sharing the encryption key between the
communicating parties) and fundamentally insecure in a group setting (
it takes just one set of ``loose lips'' to blow the security of a
single-key code that is shared by a group of individuals ).

Public key cryptography is based on a very simple, very beautiful idea
that effectively deals with both of the above issues.  The idea is
that everybody should possess two encryption keys: a public key, and a
private key.  Furthermore, the encryption method should work so that
messages encoded using person X's {\bf public} key can only be decoded
using person X's {\bf private} key.  Likewise a message encoded with a
private key should only be decodable using the corresponding public
key.  Next, everyone who wants to communicate shares their public
keys, but keeps their private key strictly hidden.  With such an
arrangement, in order to communicate with person X, the sender will
encode a message using person X's {\bf public key}.  Now everyone in
the world knows person X's public key, but it doesn't do them any
good, because it will take person X's {\bf private key} to actually
decode the message.

A system of public key cryptography can also be used to establish
identity, and to enable secure financial transactions.  Let's say that
I show up at a bank and claim that I am person X.  If I am indeed
person X, then I should be able to encode a simple message, something
to the effect of ``Hello, my name is X.'', using my secret, private
key.  Now everyone, including the bank, has access to person X's {\bf
  public} key, and so should be able to decipher the simple message,
and verify that it really was encrypted using person X's {\bf private}
key.  The point is that person X, and {\em only person X}, could have
created a message that is decodable using person X's public key.  In
this way an arrangement of public and private keys can be made to
serve as a secure authentication system (very useful for bank and
credit card transactions, as you may well imagine).

Credit for the invention of public key cryptography is typically given
to Diffie and Hillman: two computer scientists who publicized the
first public key method back in 1975.

\subsection{Modular arithmetic to the rescue}
The next issue to consider is the means by which one can implement
such a scheme of public and private key encryption.  If you've read
the preceding section on modular arithmetic, then you already have in
your possession all of the mental equipment required for such a task.

First, recall the following rule of high school algebra:
$$(a^b)^c = a^{b c}.$$
There is nothing mysterious about this
rule.  As a particular example think of the rule with $b=2$ and $c=3$.
In this instance the rule is saying that if one squares a number $a$
and then cubes the result, the final answer will be $a$ raised to the
sixth power.  Or to put it another way:
$$(a^2)^3 = (a^2)\times (a^2)\times (a^2) 
= a\times a\times a\times a\times a \times a = a^6.$$

Now in modular arithmetic, there are certain powers that do absolutely
nothing; this is the ``curious fact'' discussed in Section
\ref{sect:cc}.  For example, in Exercise \ref{ex:22sys} we saw that
$x^{21}\equiv x\mod{22}$.  Note that $21=3\times 7$.  It therefore
stands to reason that working modulo $22$, if I first raise a number
to the power $3$, and then raise the result to the power $7$, the
final result will be the number that I started with.  Eureka!  Why
don't I then use the number $3$ as a private key, the number $7$ as
the public key, and encrypt messages by raising numbers to these
powers in modulo $22$ arithmetic?  The following calculation
illustrates this.  Say you wants to send me a message that consists of
the numbers: $2, 3, 8$.  You raise each of these numbers to the power
$7$ --- that being my public key --- and so the message I receive is
$18, 9, 2$, because
$$2^7=128\equiv 18,\quad 3^7=2187\equiv9,\quad 8^7=2,097,152\equiv
2,$$
where all equivalences are $\mod{22}$.  I then take the message,
raise each number to the power $3$, and obtain
$$18^3=5832\equiv 2,\quad 9^3=729\equiv 3,\quad 2^3\equiv 8;$$
I get
back the original message!  That's all there is to it; the above
example illustrates the essentials of the RSA public key encryption
system.

\begin{exercise}
\label{ex:22rsa}
Construct a table that shows the effect of raising numbers in the
modulo $22$ system to the powers $3$ and $7$.  Inspect this table and
confirm that, indeed, the two operations undo one another.
\end{exercise}

\subsection{Some final details.}
The above discussion of the RSA algorithm has not, as yet, addressed
the following simple, but crucial question: how does one know that the
RSA public key encryption method is secure?  The fact of the matter is
that the method {\em is not secure} unless one is careful about the
way one chooses the modulus, $n$.  Consider again the example of the
preceding section: it is public knowledge that $n=22$ and that $e=7$,
but the other exponent, $f$, is not publicized.  A moment of
reflection suffices to show that knowledge of $n$ and $e$ allows one
to guess the supposedly secret value of $f$.  Indeed, knowing that
$n=22$, one also knows that $\phi=10$ (see Exercise \ref{ex:22sys}),
and hence that whatever $f$ is, $e\times f$ must be one of the
following numbers: $1, 11, 21, 31, 41, \ldots$ In other words, $f$,
whatever it is, must satisfy the equation $7\times f\equiv 1\mod{10}$.
A quick check of the numbers from $1$ to $9$ will reveal that $f=3$ is
a solution to the above equation.  The supposedly secret, private key
is thereby revealed, and the entire method of encryption rendered
useless.

The point of the above example, is that one has to be careful about
choosing the modulus $n$.  Indeed, part of the RSA methodology,
something that has not been mentioned so far, is a certain procedure
for choosing $n$ that makes the subsequent process of
encryption/decryption secure.  According to RSA, one begins by
choosing two prime numbers, call them $p$ and $q$, and sets $n=p\times
q$ (Recall that a number is called prime if it has no divisors except
for $1$ and itself.)  An added advantage of choosing $n$ in this
manner is the ease with which one can then calculate $\phi$, the
number of units modulo $n$.  
\begin{quote}
\itemhead{Fact.} If $n$ is the product of primes
$p$ and $q$, then $\phi=(p-1)(q-1)$.
\end{quote}
This formula is confirmed by the examples considered so far.  Look
back at Exercise \ref{ex:10sys}.  The modulus was $10=2\times 5$, and
the number of units was $4=(2-1)\times(5-1)$; just as predicted by the
formula.  In Exercise \ref{ex:22sys} $n$ was $22=2\times 11$, and the
modulus was $10=(2-1)\times(11-1)$, also in agreement with the
formula.

I won't prove the above formula for $\phi$ in these notes, but rather
indicate why the formula is valid by considering a particular example.
Let us consider $n=15=3\times 5$.  Arrange the numbers in the modulo
$15$ system in a three by five rectangular array like so:

\medskip
\begin{makeimage}
{\large
\begin{tabular}{c|ccccc}
   & $0$  & $1$ & $2$ & $3$ & $4$  \\
\hline
$0$ &  $0$  & $6$ & $12$ & $3$ & $9$  \\
$1$ &  $10$ & $1$ &  $7$ & $13$& $4$  \\
$2$ &  $5$  & $11$&  $2$ &  $8$&$14$  \\
\end{tabular}
}
\end{makeimage}
\medskip

What is the meaning of the above arrangement?  Every number $x$ in the modulo $15$
system can be described equally well by the following two items of
information: the value of $x$ modulo $3$, and by the value of $x$
modulo $5$.  Look, for example in row $1$, column $3$ of the above
table, and you will find the number $13$.  The reason for this
placement is that $13\equiv1\mod3$ and that $13\equiv 3\mod5$.
Similarly $12\equiv0\mod3$ and $12\equiv2\mod 5$, and therefore $12$
is located in row $0$, column $2$.

Having understood the arrangement you will see that the numbers that
are divisible by $3$ are located in row $0$, while the numbers that
are divisible by $5$ are located in column $0$.  Furthermore, the
divisors of zero in the system modulo $15$ are the numbers that are
divisible either by $3$ or by $5$ ( I leave it to you to figure out
the reason for this.)  Therefore the units of the system are going to
be all the numbers outside row $0$ and column $0$ of the above table.
In other words, the units are the numbers found in the two by four
rectangle formed by rows $1$ and $2$, and columns $1$ through $4$. No
wonder then that there is a total of $2\times 4=8$ units.

Returning to considerations of security, one needs to choose the $n$
so that the corresponding $\phi$ is enormously difficult to discover.
This is accomplished by using very large primes $p$ and $q$; it is not
unusual to take $p$ and $q$ to be a hundred digits long!  Now
multiplying two numbers a hundred digits long will produce a number
two hundred digits long --- a tedious process but one that a computer
can handle easily.  However, factoring a such a large number into its
prime components is a task that cannot be accomplished even by today's
fastest computers.  Factorization, is fundamentally a process of trial
and error, and with numbers that large, there are just too many
possibilities to check.

The upshot is that with very large $p$ and $q$, one can publicize $n$,
and the values of $p$ and $q$ will remain safely uncomputable.  Since
the knowledge of $p$ and $q$ remains secret, there is no way for an
outside agency to calculate $\phi$, and therefore no way to guess the
private key $f$, either.  The cryptosystem can then be considered to
be secure.

It is important to make one final remark regarding computations in
modular arithmetic. At first glance, the calculation of powers when
the modulus is large poses a considerable computational challenge. Say
that one wanted to calculate $48^{29} \mod{221}$. Ostensibly,

\begin{makeimage}
\vbox{
\latex{\footnotesize}
\begin{eqnarray*}
\lefteqn{48^{29} =
5,701,588,684,667,867,878,541,238,858,441,350,344,816,132,620,288}\\
&&=
2,579,904,382,202,655,148,661,194,053,593,371,196,749,381,276
\times 221 + 107.
\end{eqnarray*}}
\end{makeimage}

\noindent
In other words, this gargantuan power is equivalent to $107\mod{221}$.
This sort of brute-force calculation is beyond the capability of most
hand-held calculators.  Fortunately there is a more elegant way to do
the calculation.  Remember that $48^{29}$ is obtained by multiplying
$48$ together $29$ times.  The trick is to reduce modulo $221$ at the
intermediate stages of the calculation, rather than waiting to reduce
at the very end.  Note that $29=6\times 4+5$, and hence that
$$48^{29} = (48^6)^4\times 48^5.$$
The overall calculation can therefore be done like this (all
equivalences are modulo $221$):
$$
48^6=12,230,590,464\equiv 66,\quad
(48^6)^4 \equiv 66^4 = 18,974,736 \equiv 118,
$$
$$
48^5 = 254,803,968\equiv 29,\quad
(48^6)^4\times 48^3 \equiv 118\times 29 = 3,422 \equiv 107.
$$
There are many ways to break up this calculation.  It all depends on
how one chooses to break up the exponent $29$.
\begin{exercise}
\label{ex:largepower}
Calculate $29^{48}$ modulo $221$.
\end{exercise}

\subsection{A step by step description of the RSA algorithm.}

\begin{itemize}
\item Choose prime numbers $p$ and $q$, and set $n=p\times q$.  The
  larger the values of $p$ and $q$, the more secure the resulting
  encryption system.
\item Let $\phi$ denote the total number of units in the corresponding
  system of modular arithmetic.  Having chosen $n$ to be the product
  of primes $p$ and $q$, one has $\phi=(p-1)(q-1)$.
\item Next, one chooses two whole numbers $e$ and $f$ in such a way
  that $e\times f$ is equal to one of the critical exponents, i.e. to
  one of the following numbers: $1,1+\phi,1+2\phi,1+3\phi,\ldots$.
  This is done by first, choosing $e<\phi$ so that $e$ and $\phi$ are
  relatively prime; second, calculating the reciprocal of $e$ modulo
  $\phi$ using the modified Euclidean algorithm; and finally setting
  $f$ equal to that reciprocal.  A useful trick is to use a prime
  number less than $\phi$ as $e$.  This way one is certain that $e$
  and $\phi$ are relatively prime, and hence that $e$ is a unit modulo
  $\phi$.
\item One tells the whole world that one's public key consists of the
  modulus $n$ and the exponent $e$.  The exponent $f$ is kept a
  secret.
\item Whenever someone wants to send  a message --- and let us
  suppose that the message consists of a string of numbers modulo $n$
  --- the sender will raise each number in the message to the power
  $e$.  To decode the original message, one raises each of the
  encoded numbers to the power $f$.  
\item Conversely, to establish one's identity, one would compose a
  signature message, and raise the numbers in the message to the power
  $f$.  The receiver would then perform the authentication by raising
  the numbers in the encoded message to the power $e$, and thereby
  decode the original signature message.
\end{itemize}

\subsection{A final example.}
The present section illustrates the RSA algorithm with a simple
example.  First, one needs to choose a pair of primes --- for
instance, $p=13$ and $q=17$.  From there, $n=13\times 17=221$, and
$\phi=12\times 16=192.$

Next, one needs to choose a public key $e$ and a private key $f$ in
such a way that $e\times f\equiv 1\mod{192}$.  For this example let
$e=29$; this is a prime number, and therefore guaranteed to be a unit
modulo any $n$.  Using the modified Euclidean algorithm one calculates
that $-8\times 192+53\times 29=1$.  Hence $f=53$, because $23\times 53
\equiv 1\mod{192}$.

In order to send messages there must be a standard way to encode
letters in terms of numbers.
For the purposes of this exercise, the following encoding will be used:

\medskip
\begin{makeimage}
\halign{& # \hfil \cr
A & B & C& D& E& F& G& H& I & J & K & L & M  \cr
$1$ &
$2$ &
$3$ &
$4$ &
$5$ &
$6$ &
$7$ &
$8$ &
$9$ &
$10$ &
$11$ &
$12$ &
$13$ \cr
\noalign{\vskip 5pt}
N& O&P&Q &R&S&T&U&V&W&X&Y&Z&space \cr
$14$ &
$15$ &
$16$ &
$17$ &
$18$ &
$19$ &
$20$ &
$21$ &
$22$ &
$23$ &
$24$ &
$25$ &
$26$ &
$27$ \cr
}
\end{makeimage}
\medskip

Thus, the message ``HELLO'' corresponds to the following string of
numbers: $8$, $5$, $5$, $12$, $16$.  Encoding the message using the
public key means raising each of these numbers to the power $29$
modulo $221$.  The result is the following encoded message: $60, 122,
122, 116, 152$.  To decode the message using the private key, one
raises each of these numbers to the $53^{\mbox{rd}}$ power modulo 221,
and recovers $8, 5, 5, 12, 16$ --- the original message!

\section{Appendix. Answers to exercises.}
\itemhead{Exercise \ref{ex:add}.}  T, T, F, T.  Note that
$9+9+9=27=1+2\times13\equiv 1\mod{13}$.

\itemhead{Exercise \ref{ex:mtable}.}

\medskip
\begin{makeimage}
\large
\hskip 1em\vbox{\offinterlineskip
{\halign{\tabskip 5pt \hfil \quad$#$ & \vrule # &&\strut\hfil $#$ \cr
\times && 1 & 2 & 3 & 4  \cr
\noalign{\hrule}
1 &&  1 & 2 & 3 & 4  \cr
2 &&  2 & 4 & 1 & 3  \cr
3 &&  3 & 1 & 4 & 2  \cr
4 &&  4 & 3 & 2 & 1  \cr
}}}
\end{makeimage}

\itemhead{Exercise \ref{ex:reciprocal}.}  $3\times2\equiv1\mod 5$, and
hence $3^{-1}\equiv2\mod 5$.  Similarly, $4\times 4\equiv1\mod5$, and
therefore $4^{-1}\equiv4\mod5$.  Regarding the division problems one
has:
$4\div 3\equiv 3,\quad 3\div 4\equiv 2,\quad 3\div 3\equiv 1,\quad
1\div 3\equiv 2,$
where all equivalences are $\mod5$.

\itemhead{Exercise \ref{ex:mtable6}.}

\medskip
\begin{makeimage}
\large
\hskip 1em\vbox{\offinterlineskip
{\halign{\tabskip 5pt \hfil \quad$#$ & \vrule # &&\strut\hfil $#$ \cr
\times && 1 & 2 & 3 & 4 & 5  \cr
\noalign{\hrule}
1 &&  1 & 2 & 3 & 4 & 5 \cr
2 &&  2 & 4 & 0 & 2 & 4 \cr
3 &&  3 & 0 & 3 & 0 & 3 \cr
4 &&  4 & 2 & 0 & 4 & 2 \cr
5 &&  5 & 4 & 3 & 2 & 1 \cr
}}}
\end{makeimage}

\itemhead{Exercise \ref{ex:unit}}.
If a number $a$ is a unit then, by definition, there is a number $b$
such that $b\times a\equiv1$.  Therefore if $c\neq0$, then $a\times c$
cannot possibly be zero either, because $b\times a\times c\equiv 1\times
c\equiv c$.
On the other hand, if $a$ is a divisor of zero, then $a\times b\equiv
0$ for some non-zero $b$.   Therefore it is useless to look for a
number $c$ such that $c\times a\equiv 1$.  The reason is that if such
a $c$ were to exist then $c\times a\times b$ would be equal to
$1\times b\equiv b$.  However we know that $c\times a\times b\equiv 0$.

\itemhead{Exercise \ref{ex:10sys}}

\medskip
\begin{makeimage}
\large
\hskip 1em\vbox{\offinterlineskip
{\halign{\tabskip 5pt \hfil \quad$#$ & \vrule # &&\strut\hfil $#$ \cr
\times && 1 & 2 & 3 & 4 & 5 & 6 & 7 & 8 & 9 \cr
\noalign{\hrule}
1 &&  1 & 2 & 3 & 4 & 5 & 6 & 7 & 8 & 9 \cr
2 &&  2 & 4 & 6 & 8 & 0 & 2 & 4 & 6 & 8 \cr
3 &&  3 & 6 & 9 & 2 & 5 & 8 & 1 & 4 & 7 \cr
4 &&  4 & 8 & 2 & 6 & 0 & 4 & 8 & 2 & 6 \cr
5 &&  5 & 0 & 5 & 0 & 5 & 0 & 5 & 0 & 5 \cr
6 &&  6 & 2 & 8 & 4 & 0 & 6 & 2 & 8 & 4 \cr
7 &&  7 & 4 & 1 & 8 & 5 & 2 & 9 & 6 & 3 \cr
8 &&  8 & 6 & 4 & 2 & 0 & 8 & 6 & 4 & 2 \cr
9 &&  9 & 8 & 7 & 6 & 5 & 4 & 3 & 2 & 1 \cr
}}}
\end{makeimage}

\medskip\noindent
The divisors of zero are $2$, $4$, $6$, $8$, and $5$.  These are the
numbers that are divisible either by $2$ or by $5$.  All the other
numbers are units: $1$, $3$, $7$, $9$.

\itemhead{Exercise \ref{ex:gcd1}}
Answers: $2$, $6$, $1$, $10$.

\itemhead{Exercise \ref{ex:gcd2}}
$$1113=2\times 504+105,\quad 504=4\times 105+84,\quad 105=1\times
84+21,\quad 84=4\times 21+0.$$
Therefore $21$ is the GCD of $1113$ and
$504$.

\itemhead{Exercise \ref{ex:gcd3}}

\begin{makeimage}
\hskip 1em\vbox{\offinterlineskip
{\halign{\tabskip 2pt\hfil \quad\strut $#$  && \vrule\hfil\quad $#$ \cr
n    &p & a & b \cr
\noalign{\hrule}
1466 &  & 1 & 0  \cr
237  &6 & 0 & 1  \cr
44   &5 & 1 &-6  \cr
17   &2 &-5 & 31 \cr
10   &1 &11 &-68 \cr
7    &1 &-16& 99 \cr
3    &2 &27 &-167\cr
1    &3 &-70&433 \cr
}}}
\end{makeimage}
\medskip

\noindent
Answer: the GCD is $1$.  It is given by $1=-70\times 1466+433\times 237$.

\itemhead{Exercise \ref{ex:gcd4}}
From the preceding exercise we know that $433\times 237\equiv 1\mod
{1466}$, and hence that $237^{-1}\equiv 433\mod {1466}$.  Therefore
$$59\div 237 \equiv 59 \times 433 = 25,547 \equiv 625 \mod{1466}.$$
Checking this answer:
$$625\times 237 = 148,125 = 101\times
1466+59\equiv 59\mod{1466}.$$

\itemhead{Exercise \ref{ex:22sys}} Note that $22=2\times 11$.
Therefore, the divisors of zero in the mod $22$ system are going to be
the numbers that are divisible either by $2$ or by $11$; i.e.
$2,4,6,8,10,12,14,16,18,20$, and $11$ --- a total of eleven divisors
of zero, twelve if one includes $0$ itself.  That leaves a total of
$22-12=10$ units, i.e. $\phi=10$, and therefore the critical exponents
are going to be $p=1,11,21,31,41,51,\ldots$.

Consider a computation with $p=11$ and $x=7$.  Now
$7^{11}=1,977,326,743$.  What is this number modulo $22$? To get the
answer one must divide by $22$ and calculate the remainder.  The
division gives $89,878,488$. As for the remainder, it is
$7^{11}-89,878,488\times 22=7$, the starting number.  Try it again
with $p=21$ and $x=3$.  One gets $3^{21}=10,460,353,203$.  Dividing
the latter by $22$ one gets $475,470,600$ plus a remainder.  To
calculate the remainder one does $3^{21}-475,470,600\times 22$; the
answer is $3$, just as expected.

\itemhead{Exercise \ref{ex:22sysa}} It is evident that in the modulo
$22$ system, exponents that differ by a multiple of $10$ end up giving
the same result.  Here is why. We already noted that
$$\ldots\equiv x^{31}\equiv x^{21}\equiv
x^{11}\equiv x^1\mod{22}.$$
Multiplying through by $x$ one observes
that  
$$\ldots\equiv x^{32}\equiv x^{22}\equiv
x^{12}\equiv x^2\mod{22},$$
and of course one can keep going to conclude that
$$\ldots\equiv x^{37}\equiv x^{27}\equiv x^{17}\equiv x^7\mod{22},$$
and so on and
so forth.  In conclusion, if one starts with an exponent that is
greater than $10$ (such as $32$ for example) one can subtract a
multiple of $10$ from the exponent without affecting the answer.  In
particular, $3^{32} \equiv 3^2 \equiv 9\mod{22}$.

\itemhead{Exercise \ref{ex:22rsa}}

\medskip
\begin{makeimage}
\vbox{\offinterlineskip
{\halign{\tabskip 3pt\hfil $#$\hfil & \vrule # &&\strut\hfil $#$ \cr
x && 1 & 2 & 3 & 4 & 5 & 6 & 7 & 8 & 9 & 10 & 11 & 12 & 13 & 14 
& 15 & 16 & 17 & 18 & 19 & 20 & 21 \cr
\noalign{\hrule}
x^3 && 
1 & 8 & 5 & 20 & 15 & 18 & 13 & 6 & 3 & 10 & 11 & 12 & 19 & 16 & 9 & 4
& 7 & 2 & 17 & 14 & 21 \cr
\noalign{\hrule}
x^7  && 1 & 18 & 9 & 16 & 3 & 8 & 17 & 2 & 15 & 10 & 11 & 12 & 7 & 20
& 5 & 14 & 19 & 6 & 13 & 4 & 21 \cr
}}}
\end{makeimage}

\itemhead{Exercise \ref{ex:largepower}} The answer is $1$.  Note that
$48=6\times 4\times 2$, and hence that $29^{48}=((29^6)^4)^2$.  From
there
$$
29^6 = 594,823,321 \equiv 53,\quad
53^4 = 7,890,481 \equiv 118,
$$
$$
29^{48} = ((29^6)^4)^2 \equiv 118^2 = 13,924 \equiv 1,
$$
where all equivalences are, of course, modulo $221$.

\end{document}